\documentclass[aps,prl,reprint,groupedaddress]{revtex4-2}
\usepackage{lineno,hyperref}
\usepackage{xcolor}
\usepackage[separate-uncertainty=true,multi-part-units=single]{siunitx}
\usepackage{graphicx}

\usepackage{amsmath}
\usepackage{cleveref} 
\usepackage{breqn}
\usepackage{soul}

\crefname{equation}{Eqn.}{Eq.}
\Crefname{equation}{Equation}{Equations}
\crefname{figure}{Fig.}{Figs.}
\Crefname{figure}{Figure}{Figures.}
\crefname{section}{Sec.}{Secs.}
\crefname{table}{Table S}{Tables S}

\renewcommand*{\vec}[1]{\boldsymbol{#1}}

\newcommand*{\affmark}[1][*]{\textsuperscript{#1}}

\makeatletter
\let\cat@comma@active\@empty
\makeatother

\begin{document}


\title{Topological defect-propelled swimming of nematic colloids}


\author{Tianyi Yao\affmark[1], Žiga Kos\affmark[2,3], Yimin Luo\affmark[4], Edward B. Steager\affmark[5], Miha Ravnik\affmark[3,6] and Kathleen J. Stebe\affmark[1,*]}
\affiliation{\affmark[1]Chemical and Biomolecular Engineering, University of Pennsylvania, Philadelphia, PA 19104, USA}
\affiliation{\affmark[2]Department of Mathematics, Massachusetts Institute of Technology, Cambridge, MA 02139, USA}
\affiliation{\affmark[3]Faculty of Mathematics and Physics, University of Ljubljana, Jadranska 19, 1000 Ljubljana, Slovenia}
\affiliation{\affmark[4]Department of Chemical Engineering, University of California, Santa Barbara, CA 93106, USA}
\affiliation{\affmark[5]Mechanical Engineering and Applied Mechanics, University of Pennsylvania, Philadelphia, PA 19104, USA }
\affiliation{\affmark[6]Condensed Matter Physics Department, J. Stefan Institute, Jamova 39, 1000 Ljubljana, Slovenia}




\begin{abstract}

Non-equilibrium dynamics of topological defects can be used as a fundamental propulsion mechanism in microscopic active matter. Here, we demonstrate swimming of topological defect-propelled colloidal particles in (passive) nematic fluids through experiments and numerical simulations. Dynamic swim strokes of the topological defects are driven by colloidal rotation in an external magnetic field, causing periodic defect rearrangement which propels the particles. The swimming velocity is determined by the colloid’s angular velocity, sense of rotation and defect polarity. By controlling them we can locomote the particles along different trajectories. We demonstrate scattering -that is, the effective pair interactions - of two of our defect-propelled swimmers, which we show is highly anisotropic and depends on the microscopic structure of the defect stroke, including the local defect topology and polarity. More generally, this work aims to develop biomimetic active matter based on the underlying relevance of topology. 

\end{abstract}
\maketitle

Locomotion of colloidal swimmers, an important feature of active matter, is produced by diverse mechanisms in natural and synthetic systems\cite{lauga2009hydrodynamics,ramaswamy2010mechanics,marchetti2013hydrodynamics,lauga2016bacterial,bechinger2016active,gompper20202020}. Self-propelled swimmers convert chemical energy to generate motion with examples including bacteria and algae\cite{sokolov2007concentration,rafai2010effective,petroff2015fast} which swim by rotation of their flagella, Marangoni-stress propelled droplets\cite{maass2016swimming} which move via gradients in surface stresses, and catalytic Janus particles\cite{paxton2004catalytic,palacci2013living} which swim by phoretic motions generated by chemical reaction. Driven systems rely on external fields, including electrophoresis of charged and dielectric particles\cite{o1978electrophoretic,gangwal2008induced}, rotation\cite{tierno2008controlled, driscoll2017unstable} of magnetic particles in external fields and thermophoretic colloidal motion driven by temperature gradients \cite{piazza2008thermophoresis}. Hydrodynamic interactions, confinement, and swimmer geometry\cite{dreyfus2005microscopic,tierno2008controlled,snezhko2009self,lopez2014dynamics,molaei2014failed,lauga2016bacterial} are of central importance in such systems, affecting locomotion speed and direction, and can lead to the formation of dynamic aggregates that can be harnessed as functional structures \cite{zhang2017active,hwang2019catalytic}. 

Interactions of swimmers with their environment play important roles in determining their dynamic behavior\cite{espinosa2013fluid,qiu2014swimming}. Distinctly anisotropic environments like nematic liquid crystals (NLCs) provide means for controlling the micro-locomotion by a combination of their internal orientational order -described by a headless vector field $\vec{n}$ called the director- and their highly anisotropic viscosities. NLC can exhibit topological defects or regions of lost orientational order in the form of lines, points or even walls that strongly affect the behavior of active nematic colloids\cite{stark2001physics,lavrentovich2016active,muvsevivc2017liquid,smalyukh2018liquid,lavrentovich2021design,rajabi2021directional}. For example, bacteria align and move along the local director \cite{mushenheim2014dynamic}, causing them to accumulate at sites of splay and to be depleted from sites of bend \cite{peng2016command}. Electrophoretically driven colloids move along or perpendicular to the director field, made possible by the fluid's dielectric anisotropy \cite{lavrentovich2010nonlinear,lazo2013liquid}. Marangoni-propelled NLC droplets lose axial symmetry by flow-induced displacement of their topological defects which generates torques that result in helical trajectories \cite{kruger2016curling}. Anisotropic particles forced to rotate in NLCs display shape-dependent responses, as their companion defects become topologically unstable and rearrange with complex colloidal dynamics\cite{lapointe2004elastic,rovner2012elastic,yuan2018self,yuan2019elastic}. These latter examples show the importance of broken symmetry and defect dynamics in determining colloid behavior and indicate the general importance of dynamic effects in systems with topological defects. 

In this letter, we introduce the concept of topological defect-propelled swimming of nematic colloids. Specifically, we use ferromagnetic disks with hybrid anchoring in planar cells filled with NLC; a dipolar companion defect loop forms adjacent to the disk, pinned on the disk's sharp edges. The location of the defect defines the disk's polarity. Under continuous disk rotation, we observe periodic defect rearrangements; the disclination loop sweeps along the disk's surface, performing a `swim stroke' that propels the disk in a well-defined direction. The translation velocity $v$ is determined by the angular velocity $\omega$, the sense of rotation and the defect polarity. These nematic colloid swimmers exhibit complex interactions and form stable and unstable dimers that depend on their polarization and the topology of their defects.

\begin{figure}
\centering
\includegraphics[width=.5\textwidth]{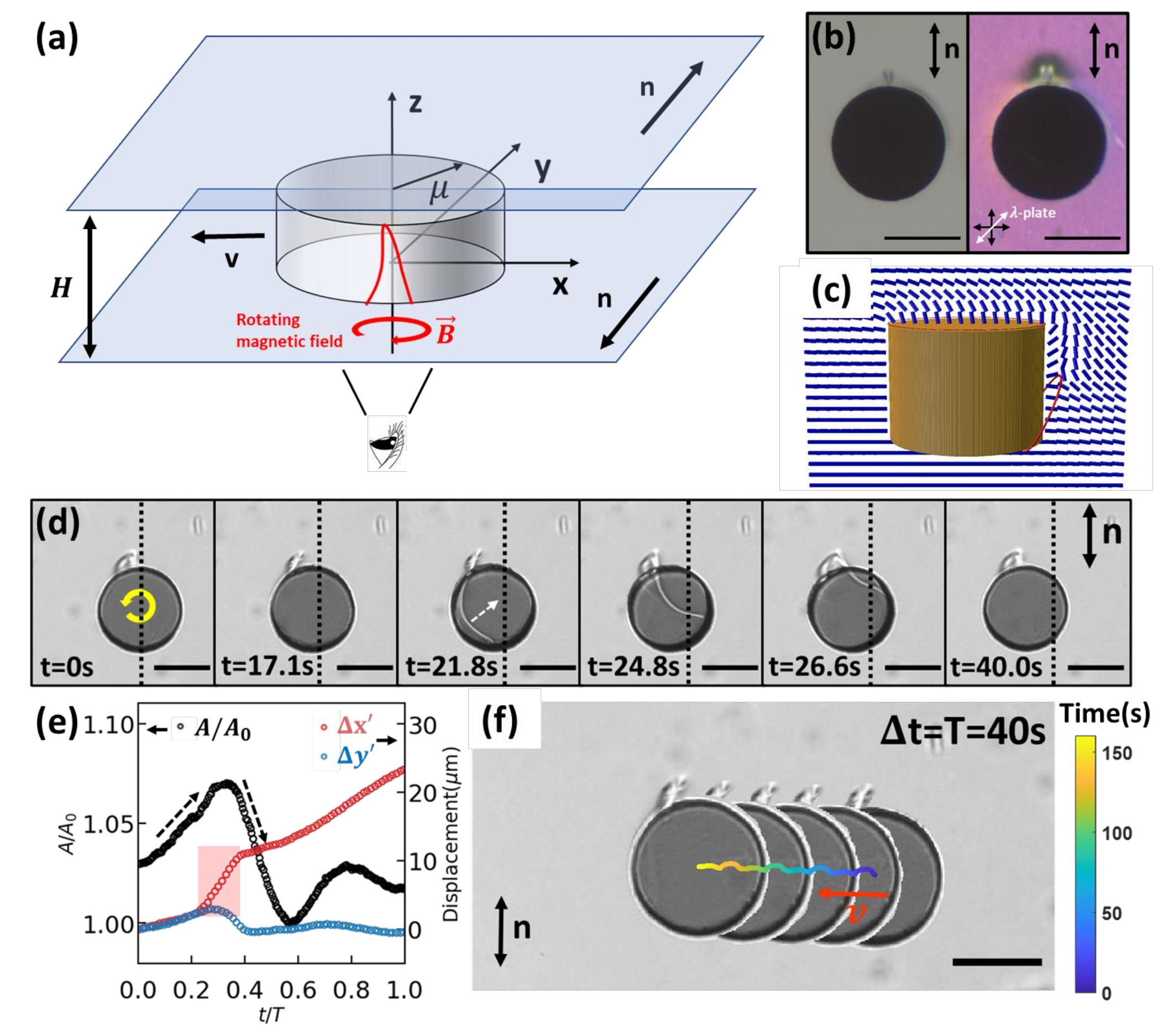}
\vspace*{-6mm}
\caption{\textbf{Far-from-equilibrium topological defect acts as a flagellum to propel nematic colloids.} (a) Scheme of experimental setup showing a magnetic disk of $2a$=\SI{75}{\micro\meter} and thickness $h$=\SI{25}{\micro\meter} sandwiched between two glass slides separated by a distance $H$ $\sim$ \SI{50}{\micro\meter} with antiparallel uniform planar anchoring. (b) Static dipolar defect configuration under bright field (left) and cross-polarization microscopy with a lambda plate (right). (c) Numerical simulation of the static dipolar configuration. (d) Time-stamped images of the swimming stroke of a disk colloid under a rotating field with period $T$=\SI{40}{\second}. The dashed line indicates the initial position of the center, and the white arrow indicates the direction of the swimming stroke. (e) Normalized projected area $A/A_0$ (black circles, left axis) and displacement (right axis) parallel (red circles) and perpendicular (blue circles) to the translation direction within one period as shown in (d). The dashed arrows indicate the tilting and flattening of the disk and the red region indicates the sweeping motion of the disclination line across the surface. The sense of rotation observed from below corresponds to clockwise rotation in (a). (f) Equal time step ($\Delta$t=$T$=\SI{40}{\second}) image showing swimming trajectory (colored curve) of the disk within \SI{160}{\second}. The red arrow indicates the velocity of the disk. Scale bars: \SI{50}{\micro\meter}.}
\vspace*{-6mm}
\label{Fig_1}
\end{figure}

Magnetic circular disk colloids were fabricated using lithographic methods followed by sputtering of a nickel film. Thereafter, homeotropic anchoring was imposed on the Ni coated surfaces and the disks were released from the substrate and dispersed in 4-cyano-4'-pentylbiphenyl (5CB) . The resulting disks have hybrid anchoring, as the surface that is not covered with Ni has degenerate planar anchoring. The colloidal suspension was introduced into the gap between two glass slides with uniform planar anchoring. Finally, the cell was placed in a programmable rotating magnetic field as shown in Fig. ~\ref{Fig_1}a (see Supplemental Material (SM) for details).

We performed numerical simulations using a Q-tensor order parameter formulation of nematodynamics to explore the stability of orientational structures and their response to the disk’s rotation. Equilibrium configurations correspond to minima of the Landau-de Gennes free energy with a surface potential describing the anchoring of nematic molecules on confining surfaces. Time evolution of the tensor order parameter is described by the Beris-Edwards model \cite{Beris}. Further information concerning the model and the numerical approach can be found in SM.

When introduced into the NLC-filled planar cell, a defect forms on the colloid with two disclination lines that connect the top and bottom faces of the disk, assuming a quadrupolar configuration (Fig.~S1 in SM). If the disk is perturbed, the defect transforms irreversibly to a static dipolar configuration that features a single loop on one side of the disk (Fig.~\ref{Fig_1}b). Numerical simulation (Fig.~\ref{Fig_1}c) reveals that this defect is a disclination loop anchored on two locations on the degenerate planar face and it extends along the side of the disk toward the homeotropic face. Fluorescence confocal polarizing microscopy corroborates the dipolar arrangement of this disclination loop (Fig.~S2 in SM). When the disk is rotated in a continuous manner, the defect undergoes a complex periodic rearrangement (Fig.~\ref{Fig_1}d) that propels the disk. The disclination line initially stretches while remaining pinned on the disk's sharp edge, storing elastic energy in the form of effective line tension of the defect core as well as elastic distortion of the NLC director field that deviates from equilibrium. These effects combined with material flow generate complex torques that cause the disk to precess, with its projected area oscillating twice in each period (Fig.~\ref{Fig_1}e). When the stored energy is high enough to depin the defect, the disclination line contracts by sweeping across the disk’s face and returns to its initial configuration. As this occurs, the disk’s tilt is reduced. The disk subsequently flattens and weakly tilts again as the defect re-forms its dipolar configuration. This swimming stroke propels the disk nearly perpendicular to the far field director in an external field at low rotation rates ($T$=\SI{40}{second} in Fig.~\ref{Fig_1}f). In each period, roughly half of the displacement occurs as the defect sweeps across the disk's face, as indicated by the red region in Fig.~\ref{Fig_1}e.

The behavior of the defect propelled nematic swimmers can be characterized in terms of the dimensionless Ericksen number which measures the ratio between viscous and nematoelastic stresses; $Er=\frac{\gamma \omega a^2}{K}$, where $\gamma$ is the rotational viscosity, $a$ is the disk radius, $K$ is the elastic constant and $\omega$ is determined by the period of the rotating field $\omega=\frac{2\pi}{T}$. We observe particle swimming at different $Er$ showing that the propulsion is affected by both the nematic elasticity and flow. The swimmer moves with Reynolds number $Re=\frac{\rho \omega a^2}{\gamma}$ $\sim$ $10^{-6}$ to $10^{-5}$, where $\rho$ is the density of 5CB. The scallop theorem dictates that non-symmetric strokes are required to swim in Newtonian fluids in creeping flow. Similarly, in the limit of small $Er$, non-symmetric defect line motion is required to achieve motion in NLC. 

\begin{figure}
\centering
\includegraphics[width=0.5\textwidth]{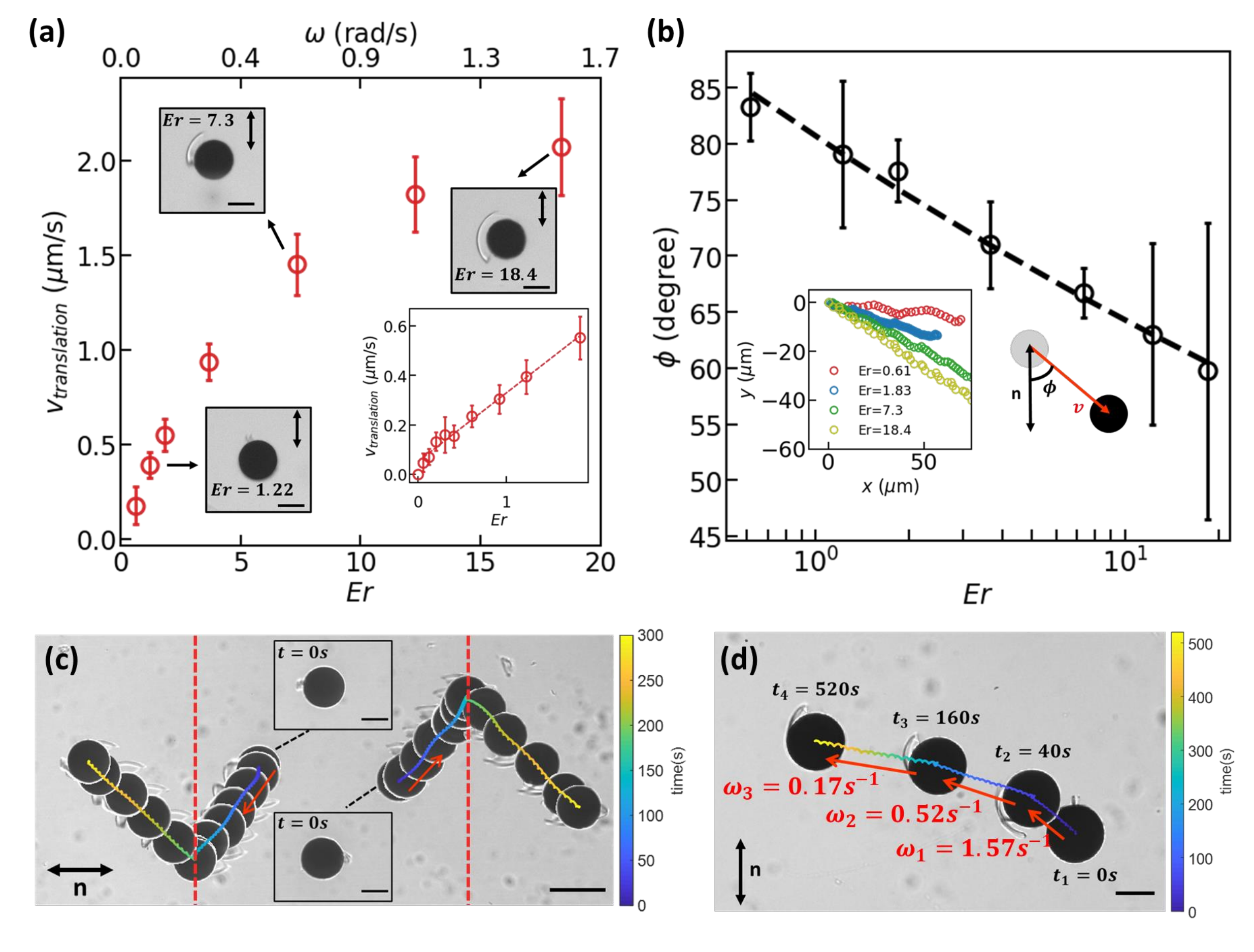}
\vspace*{-6mm}
\caption{\textbf{Effects of Er and Trajectory Planning.} (a) Translational speed and (b) direction of the defect-propelled nematic colloid as a function of $Er$. (c) Superimposed images ($\Delta t$=\SI{25}{\second}) of two colloids in same field of view with opposite polarities change swimming directions upon reversing sense of rotation of external field. The period of the external field is \SI{4}{\second} and the red dashed lines indicate the position at which the external field was switched from counter clockwise to clockwise rotation. Insets show static defect structures at $t$=\SI{0}{\second}. (d) A curved trajectory of a disk colloid in \SI{520}{\second}. Angular velocities are labeled in red for each time period  respectively. Scale bars are \SI{50}{\micro\meter} in (a), (b) and (d) and \SI{100}{\micro\meter} in (c).
}
\vspace*{-4mm}
\label{Fig_Single}
\end{figure}

A detailed examination of disk displacement and speed (see SM) over the course of the periodic defect motion indicates that translation is strongly coupled to the sweeping motion of the disclination line and the tilting motion of the disk. For slow rotation (small $Er$), the velocity $v$ is linear in $\omega$ (Fig.~\ref{Fig_Single}a) and is directed along an angle $\phi\sim$ \SI{90}{\degree} (Fig.~\ref{Fig_Single}b). The defect polarity and the disk's sense of rotation together determine the swimming direction which can be reversed by changing the disk's sense of rotation. This control of swimming direction is akin to a flagellum, which can generate ``pusher'' or ``puller" motions. As $\omega$ increases (finite $Er$), $v$ deviates from the linear relationship, attaining velocities in excess of \SI{2}{\micro\meter\per\second} for the highest frequencies probed. Moreover, the translation direction, as characterized by the angle $\phi$ in Fig.~\ref{Fig_Single}b, decreases logarithmically with $Er$. This more complex dependence of $\phi$ on $Er$ allows additional control over swimming direction; a disk colloid with given defect polarity can explore a half space by simply tuning the sense of rotation and frequency of the external field. For example, in Fig.~\ref{Fig_Single}c, two disks with opposite defect polarity move in opposite directions under the same sense of rotation. However, when their sense of rotation is reversed, rather than simply reversing their translation direction, they turn along angles that depend upon $Er$, executing V-shaped trajectories. Furthermore, the defect-propelled colloid can follow a curved path by tuning $Er$ as shown in Fig.~\ref{Fig_Single}d.

We attribute the observed swimming behavior to three possible mechanisms; detailed analysis can be found in the SM. Firstly, as the nematic configuration is periodically driven out of equilibrium at low $Er$, the disclination line elongates quasistatically, depins, and contracts at the rate determined by the balance of elasticity and viscosity. The periodic reshaping of the disclination line leads to elastic stresses acting on the disk surface and the corresponding propulsion speed can be estimated
\begin{equation}
v\sim\frac{4\pi K}{c_d\,\eta\, a},
\label{eq:v2}
\end{equation}
where $c_d$ is the drag coefficient of the disk translating between two plates and $\eta$ is the estimated average viscosity of 5CB. For experimental parameters, we obtain $\sim$ $\SI{1.3}{\micro\meter\per\second}$. Note that the disk rotation frequency is not explicitly included in Eq.~\ref{eq:v2}, but enters through the periodic forcing of the elastic deformation. Secondly, the squeezing flow in the thin film caused by the tilting of the disk exerts viscous stresses that drive translation; the resulting speed can be estimated
\begin{equation}
v \sim \frac{\Delta \eta_{gap}}{\eta} \frac{4a}{c_{d} \alpha^2 T},
\label{eq:squeezing}
\end{equation}
where $\Delta\eta_{gap}$ is the viscosity difference due to the presence of the disclination line in the thin gap
and $\alpha$ is the tilting angle of the disk with respect to the horizontal. For typical experimental values of $T$=\SI{40}{\second}, $v \sim 8.75 \frac{\Delta \eta_{gap}}{\eta} \SI{}{\micro\meter\per\second}$; taking $\frac{\Delta \eta_{gap}}{\eta} \approx 0.01-1$ for thermotropic nematics\cite{rajabi2021directional}, this expression predicts a velocity range of $0.1-10 \SI{}{\micro\meter\per\second}$, which encompasses the observed velocities. Thirdly, as $Er$ is increased, the non-linear coupling between the velocity and director field further impacts the colloid motion; the disclination loop elongates significantly adjacent to the disk (insets in Fig.~\ref{Fig_Single}a), altering the migration direction. The loop's low effective viscosity results in differences in viscous stresses on either side of the disk that bias its motion. This propulsion effect can be estimated as
\begin{equation}
v\sim \frac{\Delta \eta_{side}}{\eta} \frac{2h\omega}{c_d},
\label{eq:v1}
\end{equation}
where $\Delta\eta_{side}$ is the viscous anisotropy due to the presence of the elongated defect loop on one side of the disk. For experimental parameters at $\omega=\SI{1}{\per\second}$, this estimate yields a velocity in the range $\sim 0.02-2 \SI{}{\micro\meter\per\second}$, similar in magnitude to the other propulsion mechanisms. The elongation and subsequent instability of the disclination line along the rotating disk's edge, captured in simulation (Fig.~S7 and MovieS15 in SM) plays a prominent role in the swimming phenomenon; edge roughness may also play a role (Fig.~S3 in SM). Elasticity, broken symmetry and disclination line pinning are essential to the observed swimming mechanism. Disks rotated in 5CB in the isotropic phase fail to translate (MovieS6 in SM), and no translation was observed in experiments performed with homeotropic spherical colloids in which disclination line pinning was absent (MovieS7 in SM). Furthermore,  rotated disks with quadrupolar defects exhibit periodic disclination line pinning and release (Fig.~S8 and MovieS16 in SM) but also fail to translate (MovieS8 in SM).

\begin{figure}
\centering
\includegraphics[width=.48\textwidth]{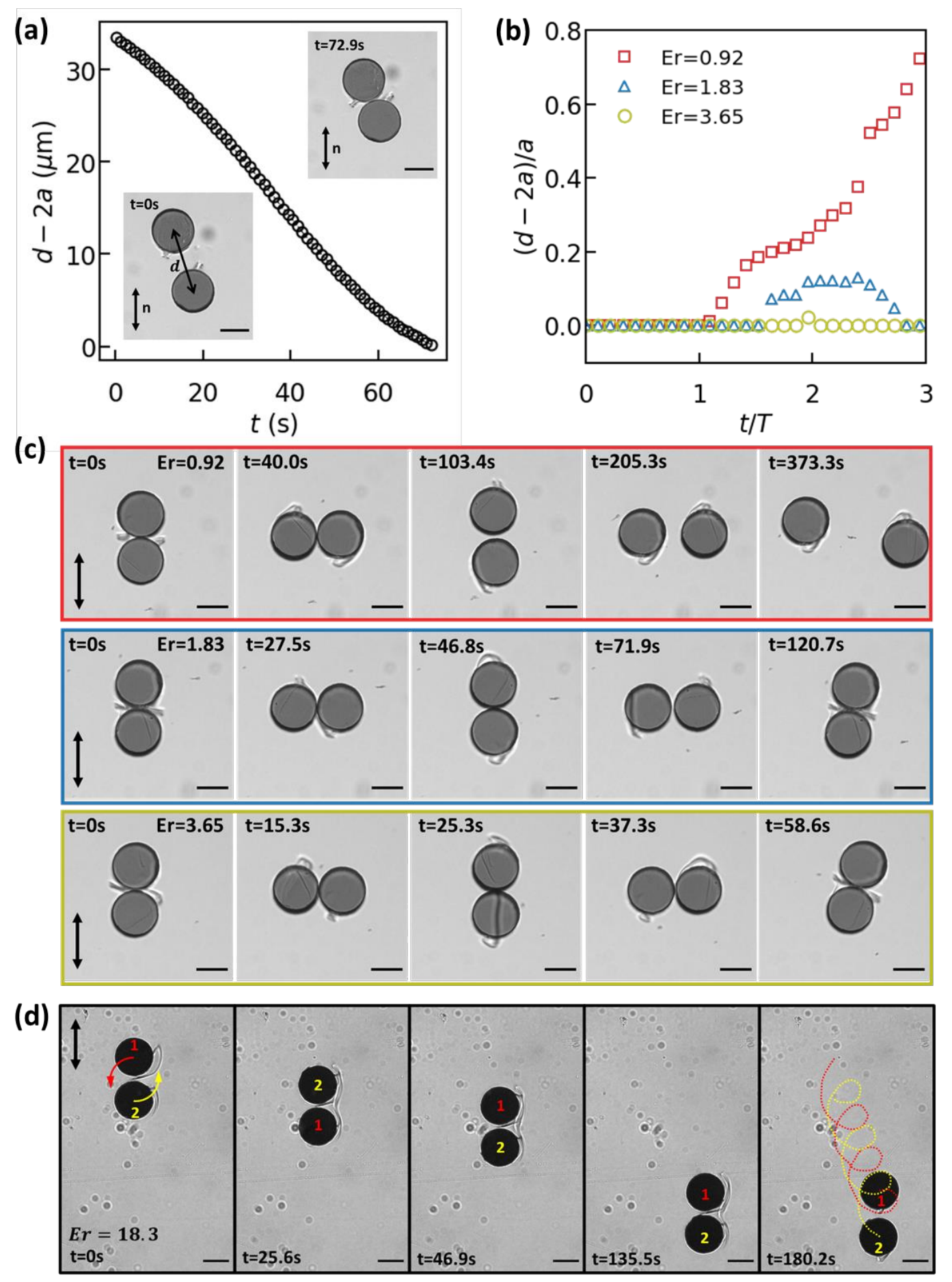}
\vspace*{-6mm}
\caption{\textbf{Pair interaction between two swimmers.} (a) Change in interparticle distance $d-2a$ between two non-swimming (i.e. no external magnetic field) colloids with opposite defect polarities. Insets show configurations at $t$=\SI{0}{\second} and \SI{72.9}{\second}. (b) Normalized interpaticle distances as a function of time of $Er$ = 0.92 (red squares), 1.83 (blue triangles) and 3.65 (yellow circles).(c) Time-stamped images of dynamic pair interactions of colloids with opposite defect polarities under rotating fields of $T$= \SI{80}{\second} (top row), \SI{40}{\second} (middle row) and \SI{20}{\second} (bottom row) respectively. (d) Time-stamped images of co-rotation and translation of dimer formed by two colloids (labeled as 1 and 2) with similar defect polarity under rotating field of $T$= \SI{4}{\second}. The red and yellow dash curves in the last frame indicated the curvilinear trajectories of colloids 1 and 2 respectively. Scale bars: \SI{50}{\micro\meter}.}
\vspace*{-4mm} 
\label{Fig_Dimer}
\end{figure}

Finally, we demonstrate pair interaction between two swimmers. Colloids at rest with opposite polarity self-assemble in a zig-zag manner (Fig.~\ref{Fig_Dimer}a) similar to their spherical counterparts with homeotropic anchoring\cite{vskarabot2007two}. Swimming introduces dynamic interactions; swimmers of opposite polarity always dimerize with the stability of the dynamic dimer depending strongly on $Er$ as shown in Figs.~\ref{Fig_Dimer}b and ~\ref{Fig_Dimer}c. At $Er\sim 0.92$, the two swimmers co-rotate as a dimer for a short period of time, separate and move away from each other (first row in Fig.~\ref{Fig_Dimer}c). For faster rotation ($Er$ $\sim1.83$), the swimmers reform a bonded dimer after a small gap is created during the co-rotation (second row in Fig.~\ref{Fig_Dimer}c). Further increasing $Er$ leads to stable dimer co-rotation as shown in the bottom row of in Fig.~\ref{Fig_Dimer}c. We hypothesize that such distinct dynamic interactions result from the interplay of repulsive interaction between unmatched boundaries and attractive defect-defect interactions which strongly depends on the $Er$ of the rotation; defects become long enough to entangle and merge above a threshold  $Er$, cementing dimers in configurations that determine their broken symmetry and swimming behavior. Dimers of opposite polarity are nearly antisymmetric, and rotate without significant translation. For example, an individual colloid swims at an average velocity of $\SI{0.55}{\micro\meter\per\second}$ while a dimer translates at $\SI{0.031}{\micro\meter\per\second}$ under the same field of $T$=\SI{40}{\second}. For disks of similar polarity, dimer formation also depends on defect elongation and topology. Disks with distinct, separate elongated defects co-migrate in the same direction without forming a dimer (MovieS12 in SM). However, above some threshold, the defects of the two disks merge to form a complex shared structure with enhanced broken symmetry; the dimers rotate and translate, as shown in Fig.~\ref{Fig_Dimer}d, with a speed of \SI{2.07}{\micro\meter\per\second}, similar to that of an individual swimmer. 

In this work, we introduce topological defect-propelled swimming of nematic colloids. We develop rotating magnetic disk colloids with complex, elongated defects which perform a `swim stroke' that drives their translation. Geometric frustration dominates for small rotation rates, and the colloid’s speed is linear in $Er$. At faster rotation rates, significant defect elongation and local changes in viscosity allow the swimming direction to be tuned for path planning. These defect-propelled swimmers exhibit far-from-equilibrium pair interactions that differ significantly from their static dipolar counterparts.

Defect-propelled swimming of nematic colloids opens opportunities for soft materials manipulation and unveils exciting questions. For example, our disks have sharp edges and hybrid anchoring conditions which generate defects that are not clearly defined by the Poincar\'{e}-Hopf or Gauss-Bonnet theorem that relate uniform anchoring to required topological charge\cite{alexander2012colloquium,beller2015shape}. Under quasistatic rotation, the defect elongates along the disk’s edge and subsequently depins from the edge; the physics that regulate these transitions and their relationship to colloid geometry are unexplored. Under finite $Er$ rotation, the dipolar defect elongates in the flow field by a dynamic instability. Our colloids also form shared, dynamically changing defects that merge and separate, subject to topological transitions whose rules are far from evident. Finally, we have reported dynamic dipolar interactions for nematic colloids, introducing a spinner colloid with significantly rate-dependent interactions. Future work will address dynamics of systems of many swimmers to probe collective phenomena including the formation of reconfigurable motile structure \cite{martinez2015magnetic,driscoll2017unstable,han2020reconfigurable}, front propagation and stability \cite{bricard2013emergence,driscoll2017unstable}.

\bibliography{ms.bib}

\end{document}


\maketitle

\section{Fabrication of ferromagnetic disk colloids and assembly of nematic liquid crystal (NLC) cell}

Circular disk colloids of diameter $2a$=\SI{75}{\micro\meter} and thickness $h$=\SI{25}{\micro\meter} were fabricated out of SU-8 photoresist following standard lithographic processes on a supporting wafer. Thereafter, a layer of nickel was sputtered onto the surface using a Lesker PVD75 DC/RF Sputterer to make the colloids ferromagnetic. Subsequently, treatment with 3 wt\% solution of N-dimethyl-n-octadecyl-3-aminopropyl-trimethoxysilyl chloride (DMOAP) imposed homeotropic anchoring condition on the disk’s Ni coated surfaces. The treated disk colloids were then released from the wafer and dispersed in 4-cyano-4'-pentylbiphenyl (5CB). Glass slides were spin-coated with polyimide (PI-2555) and rubbed with a velvet cloth along the desired direction to impose uniform planar anchoring. Two glass slides with uniform planar anchoring were assembled in an antiparallel fashion and glued together using a UV sensitive epoxy with two layers of $\sim$ \SI{15}{\micro\meter} plastic spacers in between. The resulting thickness of the cell was $\sim$ \SI{50}{\micro\meter}. Finally, a suspension of disk colloids in 5CB was introduced into the cell by capillarity in the isotropic state of 5CB. Depending on the thickness of the nickel layer, the coated disk could either appear transparent (a thin layer of nickel) or black (a thick nickel coating). While the transparent disk allowed us to visualize the swim stroke across the surface of the disk, colloids with thicker coating possess stronger magnetic moments, enabling faster rates of rotation.

\section{Controlled rotation of disk colloids}

To rotate the magnetic disk colloids, the assembled NLC cell was placed in a rotating magnetic field generated by a custom-built magnetic control system. The system consists of two orthogonal pairs of electromagnetic coils (APW Company) mounted on an aluminum supporting structure arranged around the workspace. Visual feedback is provided by a CCD camera (Point Grey Grasshopper3 Monochrome) mounted on a Zeiss inverted microscope (ZEISS Axio Vert.A1). Each coil pair was powered independently using a programmable power supply (XG 850W, Sorensen) whose outputs were controlled by a Python algorithm written in house. Sinusoidal time-dependent voltages are applied on each pair and the waveforms are separated by a $\pi$/2 phase lag in order to achieve a rotating field whose periods varied from \SI{4}{\second} to \SI{1200}{\second} for this study.

\section{Quadrupolar defect configuration around the disk colloid}

\begin{figure}[h]
\centering
\includegraphics[width=0.8\textwidth]{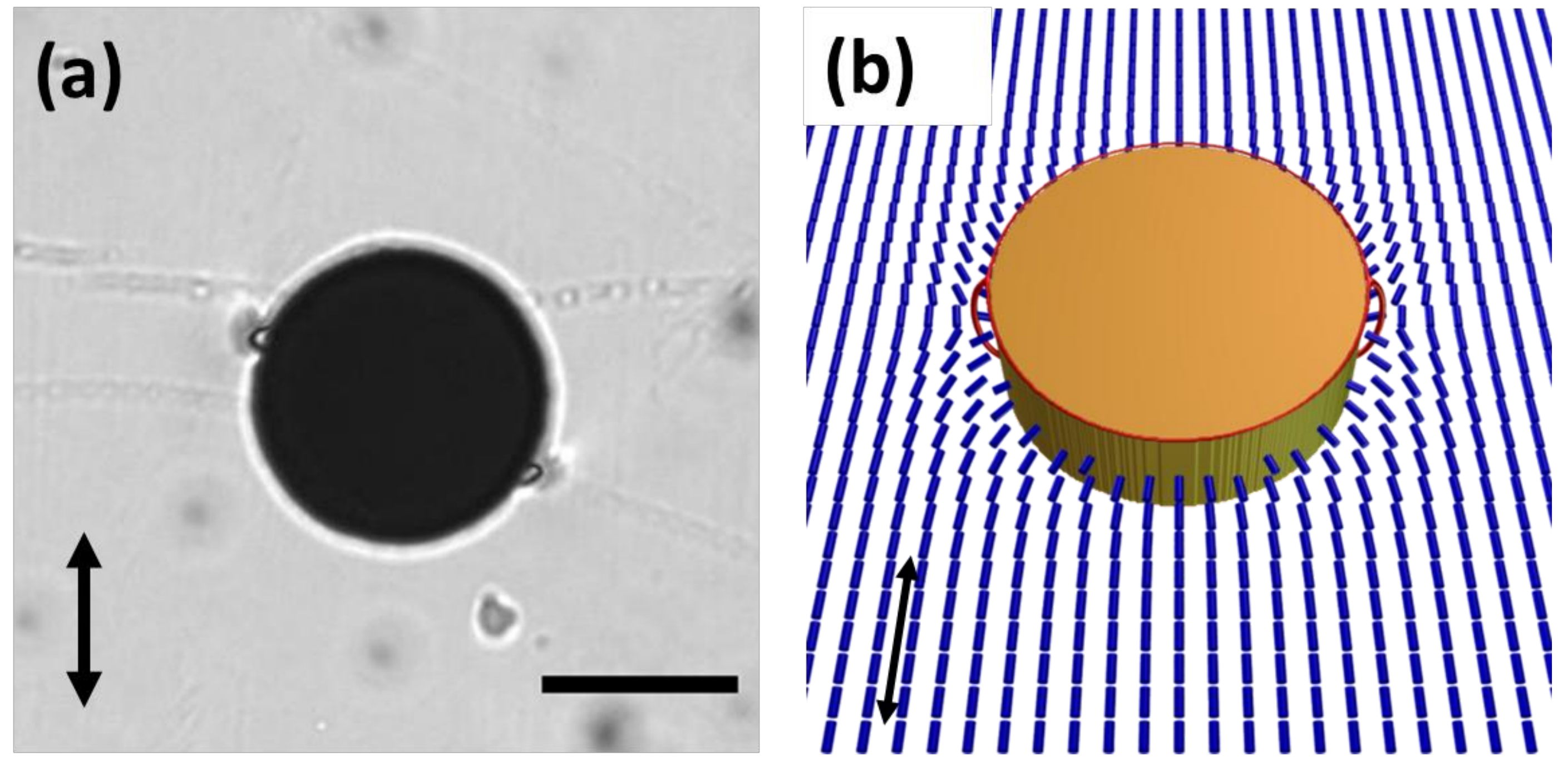}
\caption{Microscopic image (a) and numerical simulation (b) showing the quadrupolar defect configuration around a disk colloid. The double-headed arrows indicate the far-field director. Scale bar is \SI{50}{\micro\meter}.}
\label{qp}
\end{figure}

A quadrupolar defect configuration around the disk colloid is shown in Fig. ~\ref{qp}. Such a defect configuration sometimes forms after quenching the NLC to the nematic phase from the isotropic phase; this configuration is metastable, and readily transforms into the dipolar configuration under a rotating external magnetic field as shown in supporting movie S1. 

\section{Characterization of static dipolar defect using fluorescent confocal polarizing microscopy (FCPM)}

\begin{figure}[h]
\centering
\includegraphics[width=0.8\textwidth]{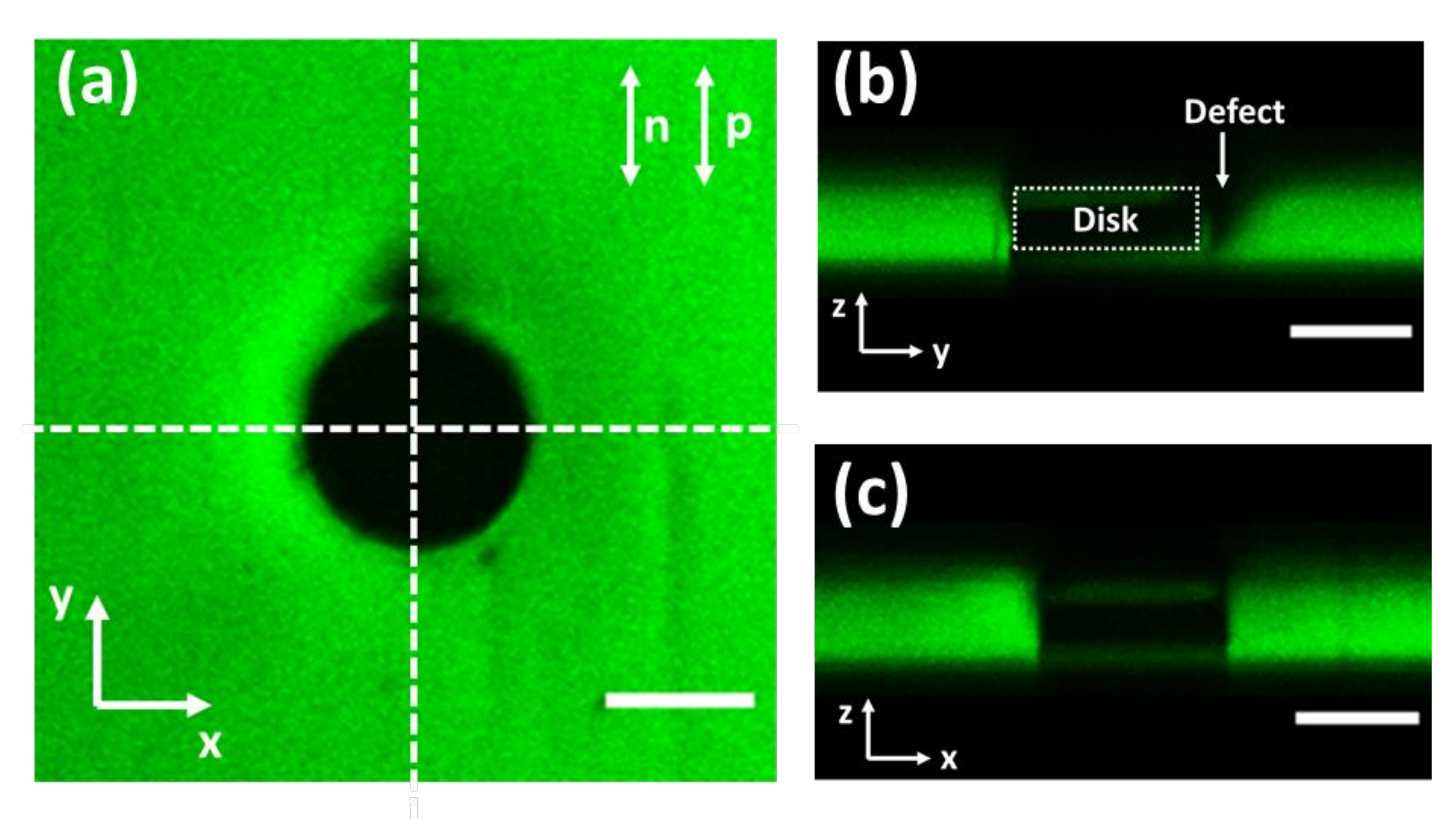}
\caption{Fluorescent confocal polarizing microscopy (FCPM) images of a disk colloid with dipolar defect shown in (a) xy-plane, (b) yz-plane and (c) xz-plane. Scale bars are \SI{50}{\micro\meter}.}
\label{confocal}
\end{figure}

The configuration and location of the defect in a dipolar configuration around a disk colloid along the vertical z-axis was determined by FCPM\cite{SMALYUKH200188}. The NLC, 5CB, was doped with an anisotropic dye N,N`-Bis(2,5-di-tert-butylphenyl)-3,4,9,10-preylenedicarboximide (BTBP; Sigma–Aldrich) at 0.01 wt\%. At such low concentration, dye molecules co-align with the NLC molecules while preserving the properties of 5CB and fluoresce when aligned parallel to the polarization direction of the excitation light. FCPM images of the disk colloid with a dipolar defect in a planar NLC cell (Fig. ~\ref{confocal}) were obtained using an inverted IX81 Olympus microscope equipped with an FV300 Olympus confocal scan box. A polarizer was placed between the sample and the objective to rotate the polarization of the scanning laser. As shown in Fig. ~\ref{confocal} and supplemental movie S2, a dipolar defect forms on top of the disk aligned with the far-field director. In addition, FCMP scanning along the z-axis (supporting movie S2) and the cross section of the yz plane (Fig. ~\ref{confocal}b) indicate that this dipolar defect is in the form of a disclination loop anchored on the top and bottom edge of the disk and the cross-sectional area of the domain containing the loop is changing along the z-axis. This configuration is in agreement with the results from numerical simulations shown in Fig 1(c) in the main text.

\section{Rotation of disk colloids in the isotropic phase of 5CB}
\label{sec:rotation_isotropic}

To examine the importance of NLC elasticity in the observed translation of the rotated disk, the same colloids were rotated in the isotropic phase of 5CB. A planar cell filled with the disk colloid suspension was sealed on all sides using epoxy resin and submerged in a hot water bath with temperature higher than the phase transition temperature of 5CB. The 5CB in the sealed cell was melted into the isotropic state and the water bath ensured that any possible drift from temperature gradients was minimized. An external field was applied to rotate the disk in the isotropic phase of 5CB. In comparison to disks rotated in the nematic phase, the translational velocity was attenuated by more than an order of magnitude (see supplemental video S6). For example, the apparent velocities of disk colloids rotating in isotropic and nematic phase of 5CB are \SI{0.025}{\micro\meter\per\second} and \SI{0.94}{\micro\meter\per\second} respectively under the same external field with period $T$= \SI{20}{\second}. Notably, the velocity is decreased even though the viscosity is reduced; the viscosity of 5CB decreases with temperature, and is lower in the isotropic phase than in the nematic phase.

\section{Characterization of surface roughness of the disk colloids using atomic force microscopy (AFM)}
\label{sec:afm}

\begin{figure}[h]
\centering
\includegraphics[width=1\textwidth]{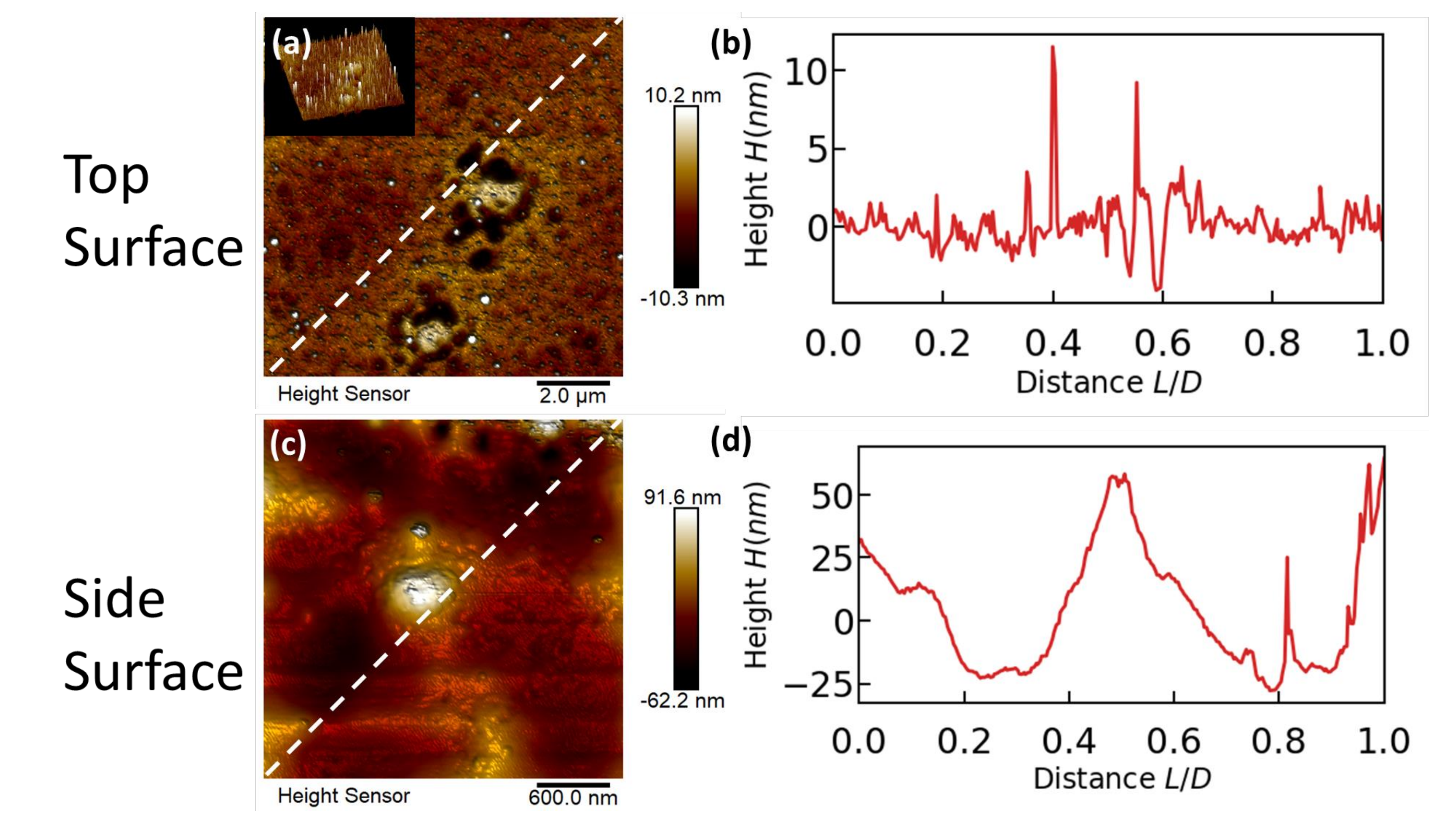}
\caption{Roughness of the disk colloid’s top and side surfaces: (a) and (c) AFM images of a \SI{10}{\micro\meter} by \SI{10}{\micro\meter} area on the top surface of the disk and a \SI{3}{\micro\meter} by \SI{3}{\micro\meter} area on the side surface of the disk, respectively. Scale bars are \SI{2}{\micro\meter} and \SI{600}{\nm}. (b) and (d) indicate the height profile along the white dash diagonals in (a) and (c). Note difference in scale in (b) and (d).}
\label{afm}
\end{figure}

The surface roughness of the top and side surfaces of the disk colloid after Ni deposition was characterized using a Bruker Icon AFM in standard tapping mode. For the flat surface, the colloid’s surface roughness was measured while still attached to a silicon substrate. Characterization of a \SI{10}{\micro\meter} by \SI{10}{\micro\meter} area gave a root-mean-square roughness $R_q$=\SI{3.29}{\nm} with a peak value of \SI{107}{\nm}, indicating that the top surface is nanoscopically smooth with a few isolated rough sites in the hundred nanometer range as shown in Figs. ~\ref{afm}(a) and ~\ref{afm}(b). The roughness of the side surface was obtained by placing the disk on its side, adhered to a planar support. The root-mean-square roughness $R_q$ of a \SI{3}{\micro\meter} by \SI{3}{\micro\meter} scanning area (Figs. ~\ref{afm}(c) and ~\ref{afm}(d)) was \SI{19.4}{\nm} with a peak value of \SI{213}{\nm}. The side surface has greater roughness than the top surface. We attribute such surface roughness to non-uniform Ni deposition and typical resolutions achieved in 2D UV photolithography. These rough sides may facilitate the pinning of the disclination line. It is interesting to note that the roughness on the sides and edges are similar in size to the defect core $\sim\SI{10}{\nm}$.

\section{Controlled rotation of spherical colloids in the nematic phase of 5CB}

To demonstrate the importance of defect line pinning in the observed translation, $2a$=\SI{8.74}{\micro\meter} ferromagnetic spherical colloids (Spherotech, Inc.) were treated with DMOAP to impose homeotropic anchoring on the surface before being dispersed in 5CB. The colloidal suspension was introduced into a planar cell of thickness $H$ $\sim$ \SI{30}{\micro\meter} and the cell was placed under an external rotating magnetic field with period $T$= \SI{6}{\second}. Spherical colloids rotated with the field with their hedgehog defect oscillating around the initial equilibrium as shown in supporting movie S7. During this process, no translation was observed. Notably, the defect in this example was not pinned and did not stretch as the colloid rotated. Rather, the companion defect moved freely above the colloid.

\section{Co-migration of two colloids with same defect polarity without dimerization}

Two rotating disk colloids with the same polarity co-migrated without dimerization or defect entanglement below a threshold Ericksen number. The pair co-migrated in a head-to-tail fashion and moved along a well-defined direction similar to the direction selected by individual swimmers as shown in supporting movie S12. However, the swimming velocities of the co-migrating pair were much lower than individual colloids under the same external field.


\section{Derivation and scaling analysis on the propulsion mechanisms}
\label{sec:mechanism}

Disk propulsion is linked to the anisotropic order within the nematic liquid crystal, as shown in experiments in Section~\ref{sec:rotation_isotropic}. In this section, we discuss the relevant mechanisms of propulsion and their scaling with the disk size, viscoelastic properties of the nematic, and disk rotation frequency. The propulsion mechanisms below are roughly on the same order of magnitude and likely all contribute to the final speed of propulsion. This is not surprising, as all of the mechanisms are due to pushing the nematic configuration out of equilibrium.

\subsection{Elastic forces}
In a nematic medium, elastic stress is exerted on confined objects, which is described by the Ericksen stress tensor~\cite{KlemanM}. The elastic stress aims to displace the objects in order to minimize the nematic elastic free energy. If we assume that the length scale of the director distortion is of the order of disk radius $a$, the Ericksen stress tensor scales as $\sigma_{\text{Er}}\sim K/a^2$, where $K$ is the nematic elastic constant. The propulsion force is calculated by integrating the Ericksen stress over the disk surface, which scales as $4\pi a^2$, assuming that the disk thickness is of similar size as the radius. The obtained scaling for the propulsion force is therefore of the order of the elastic constant $F_{\text{prop}}\sim K$.

Another approach towards estimating the elastic forces on the disk is through the concept of defect line tension. Geometric frustration due to rotation of a disk particle with a hybrid anchoring pattern generates a periodic reshaping of a pinned defect line. During this periodic swimming stroke, the pinned defect line pulls on the colloidal particle and they collectively propel across the nematic liquid crystal. A defect line pinned to the disk surfaces acts upon it with a force of $\pi K/4\log\left( a/a_{\text{core}} \right)$~\cite{KlemanM}, where we have again used the radius $a$ as a measure of the length scale and $a_{\text{core}}$ is the size of defect line cores (for 5CB $a_{\text{core}}$ is of the order of a few nm). As the logarithmic factor changes very slowly with $a$ and the obtained result for the experimental radius is similar in magnitude to the $F_{\text{prop}}$ above, we shall use the scaling of $F_{\text{prop}}\sim K$ in our analysis. Furthermore, in equilibrium, the elastic forces cancel themselves out and there is no net force on the disk. The above scaling of the propulsion force is valid at disk high rotation rates, where the nematic field is far from equilibrium. 

From the propulsion force $F_{\text{prop}}$, we can write the translation velocity
\begin{equation}
v_{\text{translation}}=\frac{F_{\text{prop}}}{c_d\,\eta\, a} \sim \frac{4\pi K}{c_d\,\eta\, a},
\label{eq_v_elastic}
\end{equation}
where $c_d$ is the linear drag coefficient for the disk and $\eta$ is the effective viscosity of 5CB, calculated as an average of Miesowicz viscosities. For a disk translating in a confined environment between two parallel plates, the drag coefficient is estimated from lubrication theory as $c_d \sim c_{d,bottom} + c_{d,top} \sim \frac{4a}{h_0} + \frac{4a}{h_1}$, where $h_0$ and $h_1$ are the gap thickness between the disk surface and the bottom plate and the top plate, respectively. For a typical cell we used in the experiment, FCPM reveals that $h_0 \sim \frac{1}{2} h_1 \sim \SI{8.6}{\micro\meter}$. By defining a dimensionless parameter $\varepsilon = \frac{h_0}{2a}$, a drag coefficient $c_d= \frac{3}{\varepsilon}$ will be used for further analysis. For the relevant parameter values of $K=\SI{6.5}{\pico\newton}$~\cite{KlemanM}, $a=\SI{37.5}{\micro\meter}$, $\eta=\SI{0.064}{\pascal\second}$~\cite{KlemanM}, we obtain the value of translational velocity of $v_{\text{translation}}\sim\SI{1.3}{\micro\meter\per\second}$, which is of similar order of magnitude as the experimental results at high frequency of disk rotation.

\subsection{Hydrodynamic force due to squeezing flow}

During the defect rearrangement process, the disk periodically tilts and flattens as the disclination line elongates and sweeps across the disk surface. This squeezing flow results in a hydrodynamic force in the thin gaps between the disk and the bounding plates. On one side of the disk, where the defect performing the swimming stroke is absent, the disk does not experience a net force due to the scallop theorem since the tilting and flattening are completely reversible. However, the defect sweeps over the other side of the disk, causing it to tilt, and is absent during the flattening process. This viscosity difference avoids the constraints of the scallop theorem, and results in a net hydrodynamic force on the disk. To estimate the propulsion velocity from this force, we assume the viscosities of the fluid in the thin gap when the disclination line is present and absent are $\eta_{1,gap}$ and $\eta_{2,gap}$, respectively. Under thin-film approximation, the net hydrodynamic force in the horizontal direction  due to the squeezing flow as the disk tilts and flattens is $F_x \sim \frac{4a^2 \omega \Delta\eta_{gap}}{\alpha^3} \sim \frac{4a^2 \Delta\eta_{gap}}{\alpha^2 T}$, where $\Delta \eta_{gap} = \eta_{1,gap} -\eta_{2,gap}$, $\alpha$ is the tilting angle with respect to the horizontal direction and $T$ is the period of the process. This force is balanced by a drag force along the x direction $F_D \sim c_d \eta a v_{\text{translation}} $. Equating $F_x$ and $F_D$ gives the translational velocity of the disk 
\begin{equation}
v_{\text{translation}} \sim \frac{\Delta \eta_{gap}}{\eta} \frac{4a}{c_{d} \alpha^2 T}.
\end{equation}
Using typical experimental values for $T$=\SI{40}{\second}, $v \sim 8.75 \frac{\Delta \eta_{gap}}{\eta}\, \SI{}{\micro\meter\per\second}$; taking $\frac{\Delta\eta_{gap}}{\eta} \approx 0.01-1$ for thermotropic nematics\cite{rajabi2021directional}, this predicts a velocity range $\sim 0.1-10\, \SI{}{\micro\meter\per\second}$ which contains observed velocities.

\subsection{Nonhomogeneous viscosity on either side of the colloid}

As the disk rotates in the nematic fluid, it experiences a viscous force that is dependent on the structure of the nematic field. The side of the disk with the nematic defect present is in a region of different (and in principle lower) effective viscosity than the opposite side of the disk. Such a difference in viscosities can explain the propulsion of spinning disks in the direction that is observed in experiments (Fig.~1). To estimate the magnitude of propulsion velocity due to different viscosity regions, we assume that the effective viscosity on one side of the disk equals $\eta_{1,side}$ and on the other side $\eta_{2,side}$. The shear at the disk side-wall is estimated to $v_{\text{side-wall}}/a=\omega$. The shear force density on the side-wall therefore equals $\mathrm{d}F_1/\mathrm{d}S=\eta_{1,side}\omega$ in the region of $\eta_{1,side}$. Integrating over $\mathrm{d}S=ah\,\mathrm{d}\phi$, where $h$ is the disk thickness, we obtain $F_1=\int_0^\pi\mathrm{d}\phi\,ah\,\omega\eta_1\sin\phi=2a h \omega\eta_{1,side}$. The $\sin\phi$ term is included as we are interested only in one component of the force. The net force on the disk can be written as $F_{\text{prop}}\sim F_1-F_2=2ah\omega\Delta\eta_{side}$, where $\Delta\eta_{side}=\eta_{1,side}-\eta_{2.side}$. The net force generates a disk translational velocity of
\begin{equation}
v_{\text{translation}}\sim \frac{2h\omega\Delta\eta_{side}}{c_d\,\eta}.
\end{equation}
Using the same parameters as for Eq.~\ref{eq_v_elastic}, disk thickness $h=\SI{25}{\micro\meter}$, experimental rotational velocity from Fig.~2 of $\omega=\SI{1.0}{\per\second}$, and range of viscous anisotropy $\frac{\Delta\eta_{side}}{\eta}$ mentioned above, we obtain the approximate value for the translational velocity of $v_{\text{translation}}$ in the range $\sim 0.02-2 \mu m s^{-1}$ which is similar in magnitude to the other propulsion mechanisms.


\section{Speed variation within one period}

\begin{figure}[h]
\centering
\includegraphics[width=0.65\textwidth]{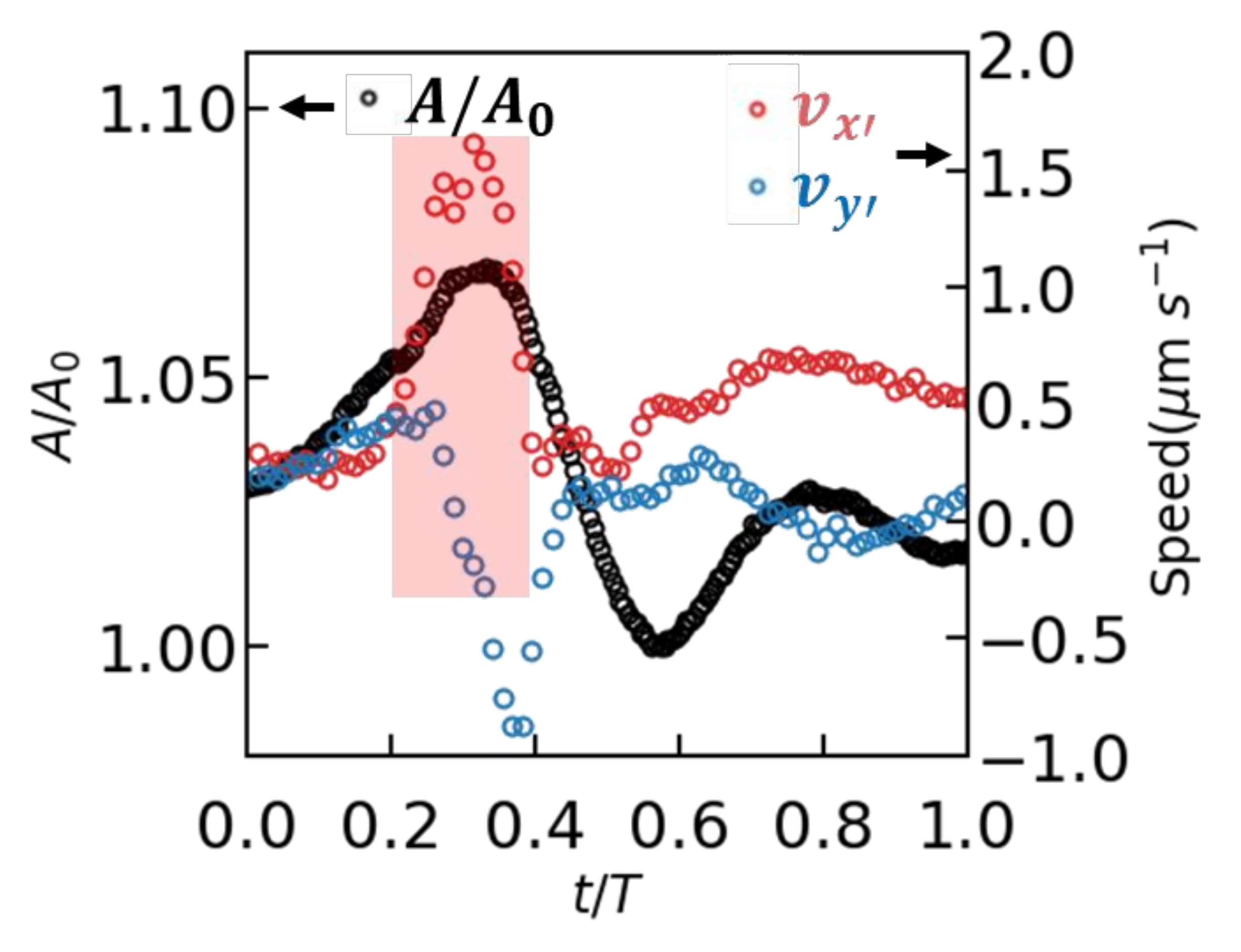}
\caption{Speed variation within one period: Normalized projected area $A/A_0$ (black circles, left axis) and swimming speed (right axis) parallel (red circles) and perpendicular (blue circles) to the translation direction within one period corresponding to Fig.~1(e) in the main text.}
\label{speed}
\end{figure}

The sweeping motion of the disclination and the processing motion of the disk play a significant role in propelling the colloid swimmer. As shown in Fig.~1(e) in the main text, roughly half of the displacement occurs in less than one fifth of the period during which the disclination line sweeps across the surface of the disk in the thin gap and the disk tilts and flattens. This coupling is more clear if we further look at the speed variation of the swimmer within one period as shown in Fig.~\ref{speed}. A sharp peak of swimming speed along the translation direction $v_{x'}$ (red circles) is observed during the sweeping motion of the disclination line (red region) and is clearly coupled with the tilting and flattening of the disk (black circle).

\section{Details on the numerical simulations}

We performed numerical simulations using a Q-tensor order parameter description of nematodynamics. The director field $\vec{n}$ is obtained as a main eigenvector of the Q-tensor and the degree of order $S$ as its main eigenvalue. Equilibrium structures correspond to a minimum of the free energy
\begin{equation}
\begin{split}
F=&\int_{\text{bulk}}\mathrm{d}V\left[ \frac{A}{2}Q_{ij}Q_{ji}+\frac{B}{3}Q_{ij}Q_{jk}Q_{ki}+\frac{C}{4}\left(Q_{ij}Q_{ji}\right)^2 +\frac{L}{2}(\partial_k Q_{ij})(\partial_k Q_{ij}) \right] \\
+ &\int_{\text{disk bottom surf.}}\mathrm{d}S\, W \left(\tilde{Q}_{ij}-\tilde{Q}_{ij}^\perp\right)^2,
\label{eq:F}
\end{split}
\end{equation}
which is calculated from the bulk and the surface contributions. $A$, $B$, and $C$ are nematic phase parameters and dictate the degree of order in the equilibrium homogeneous director field $S_{\text{eq}}$. $L$ is the tensorial elastic constant and is directly proportional to the director elastic constant $K$. The surface integral is performed only over the disk bottom surface, where the anchoring of nematic molecules is planar degenerate. On the disk side-walls, disk top surface and the cell's top and bottom boundaries, the director field is fixed. The planar-degenerate surface is modeled by a Fournier-Galatola potential, where $\tilde{Q}_{ij}=Q_{ij}+\frac{S_{\text{eq}}}{2}\delta_{ij}$, $\tilde{Q}_{ij}^\perp=(\delta_{ik}-\nu_i\nu_k)\tilde{Q}_{kl}(\delta_{lj}-\nu_l\nu_j)$, and $\vec{\nu}$ is the surface normal.

Q-tensor dynamics is described by the Beris-Edwards model~\cite{KlemanM}
\begin{equation}
\dot{Q}_{ij}=\Gamma H_{ij}+S_{ij},
\label{eq:Qdot}
\end{equation}
where $H_{ij}$ is the molecular field driving the nematic orientation towards a free energy minimum
\begin{equation}
H_{ij}=-\frac{1}{2}\left( \frac{\delta F}{\delta Q_{ij}} + \frac{\delta F}{\delta Q_{ji}} \right)+\frac{1}{3}\frac{\delta F}{\delta Q_{kk}}\delta_{ij}
\end{equation}
and $\Gamma$ is the rotational viscosity parameter. $S_{ij}$ describes the nematic response to flow gradients
\begin{equation}
S_{ij}=\left( \zeta A_{ik}-\Omega_{ik} \right) \left( Q_{kj}-\frac{\delta_{kj}}{3} \right)+ \left( Q_{ik}-\frac{\delta_{ik}}{3} \right)\left( \zeta A_{kj}-\Omega_{kj} \right) -2\zeta\left( Q_{ij}+\frac{\delta_{ij}}{3}\right)Q_{kl}\partial_l v_k,
\end{equation}
where $\vec{v}$ is the flow field, $A_{ij}=(\partial_iv_j+\partial_jv_i)/2$, $\Omega_{ij}=(\partial_iv_j-\partial_jv_i)/2$, and $\zeta$ is the nematic alignment parameter. 
On the planar degenerate surface, the Q-tensor follows the dynamics of
\begin{equation}
\dot{Q}_{ij}^{\text{surf}}=\Gamma_{\text{surf}}\left[ \frac{1}{2}(H_{ij}^{\text{surf}}+H_{ji}^{\text{surf}})-\frac{1}{3}\delta_{ij}H_{kk}^{\text{surf}} \right]
\end{equation}
where $\Gamma_{\text{surf}}$ is the surface rotational viscosity parameter, and
\begin{equation}
H_{ij}^{\text{surf}}=-\frac{\partial f_{\text{vol}}}{\partial(\partial_k Q_{ij})}\nu_k -\frac{\partial f_{\text{surf}}}{\partial Q_{ij}},
\end{equation}
is the surface molecular field, calculated from the bulk and surface free energy density given by Eq.~\ref{eq:F}.
Simulations where the disk does not rotate, were solved using an approximation of no flow. For simulations with rotating disk, the flow field was calculated by a lattice Boltzmann method with a moving boundary condition, and the resulting stationary flow field was used in Eq.~\ref{eq:Qdot}.

Simulations were performed using a finite difference method to solve Eq.~\ref{eq:Qdot}. The simulation in Fig.~\ref{SI_fig2} was obtained on a $580\times 580\times 140$ mesh size with disk radius $a=105\,\Delta x$ and disk height $h=70\,\Delta x$ ($\Delta x$ is the mesh resolution), while other simulations were performed on a $380\times 380\times 280$ mesh with $a=95\,\Delta x$ and $h=140\,\Delta x$. A plane with a no-slip velocity boundary condition and a fixed planar director field is used at the top and at the bottom of the simulation box. Periodic boundary conditions are used in the lateral directions of the numerical simulation box. Mesh resolution is set as $\Delta x=1.5\xi_{\text{N}}=1.5\sqrt{L/(A+BS_{\text{eq}}+\tfrac{9}{2}CS_{\text{eq}}^2)}$, where $\xi_{\text{N}}$ is the nematic correlation length that sets the size of the defect cores. The following values of the model parameters are used: $\zeta=1$, $B/A=12.3$, $C/A=-10.1$, $W=0.5\,L/\Delta x$, $\Gamma_{\text{surf}}=\Gamma/\Delta x$ (unless otherwise specified), and a timestep of $0.1(\Delta x)^2/(\Gamma L)$.
The results of the simulations are expressed using the mesh resolution $\Delta x$, rotational viscosity parameter $\Gamma$, and elastic constant $L$.

\begin{figure}[h]
\centering
\includegraphics[width=\textwidth]{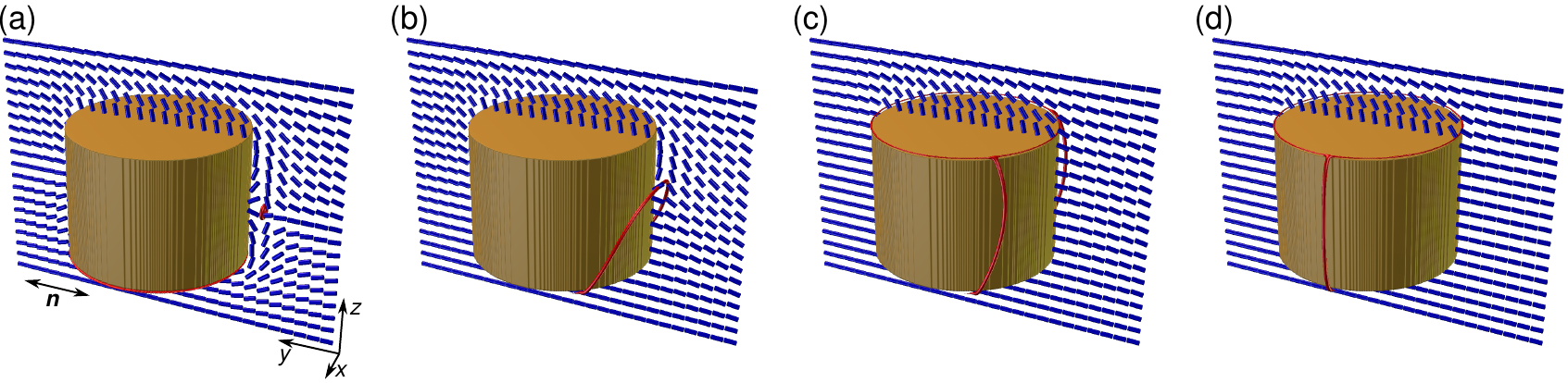}
\caption{Defect configurations in numerical simulations. Depending on the boundary condition on the disk surface, nematic configuration in simulations can show various dipolar or quadrupolar solutions. In all panels, the disk bottom surface is perfectly planar degenerate and the top surface is homeotropic. The disk is positioned in a cell with the director orientation along $y$ axis on the top and the bottom plane. (a) On the side-wall of the disk a splay-like director field is fixed, going from $-z$ to $+z$ orientation between the bottom and the top surface. Such a boundary condition stabilizes a point defect in the form of a small loop along the director far-field axis. (b) If the splay-like boundary condition is enforced only on the top half of the disk's side-wall, a defect line is the stable dipolar configuration. The defect line is pinned at two points to the bottom edge of the disk, and its cross-section shows a half-integer profile. (c) If the boundary condition on the whole side-wall is homeotropic, the nematic field evolves from the initial dipolar ansatz into a solution with two defect lines, each pinned to the top and the bottom edge. This dipolar solution is unstable and in time evolves into the quadrupolar field (d). The time evolution from dipolar to the quadrupolar field is shown in Fig.~\ref{SI_fig2}.}
\label{SI_fig1}
\end{figure}

\begin{figure}[h]
\centering
\includegraphics[width=\textwidth]{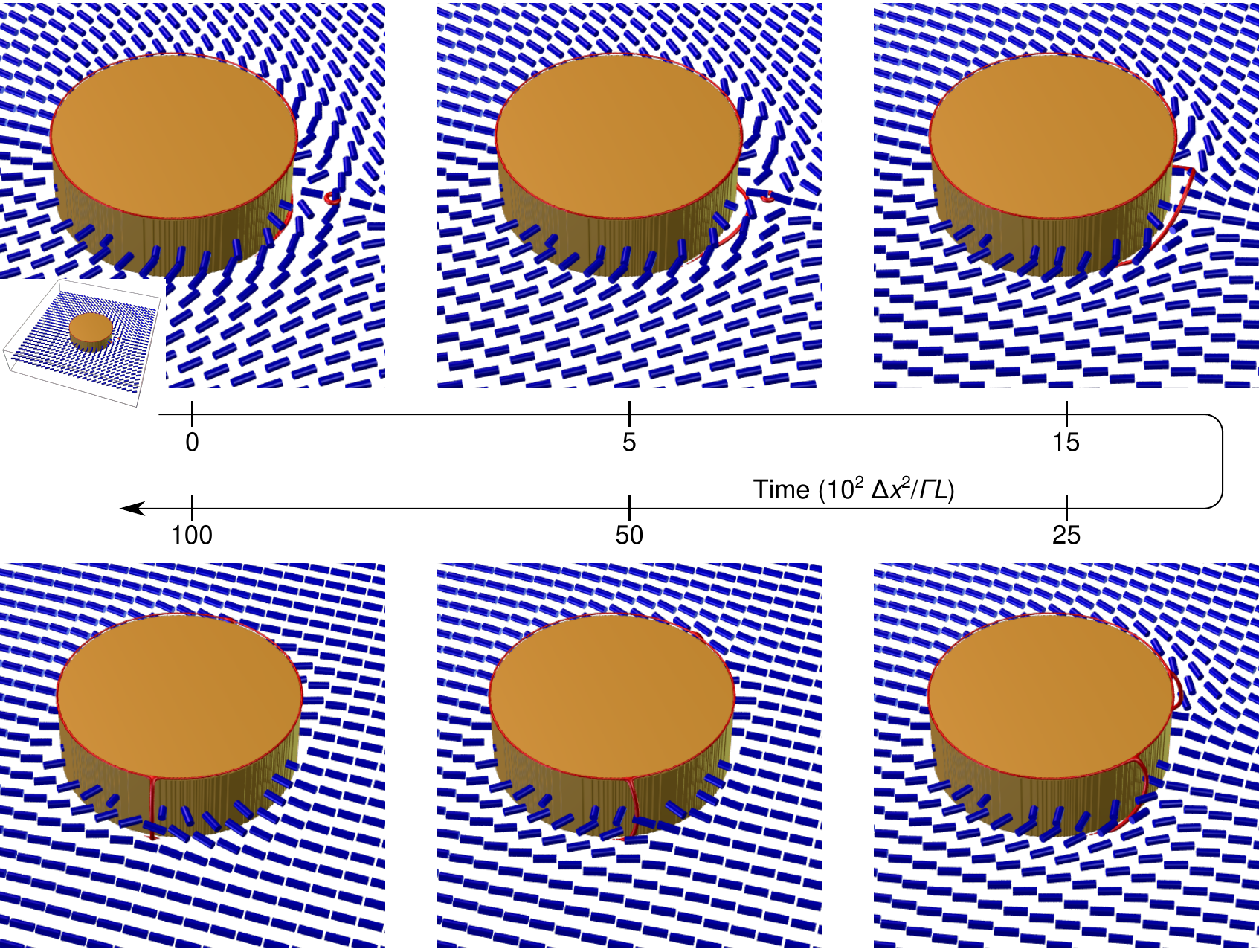}
\caption{Time evolution from the dipolar to the quadrupolar configuration. The initial configuration is a dipolar director field with a point defect not in contact with the disk surface. Inset shows the disk position and dimension inside the numerical simulation box. In time, the point defect comes in close proximity to the defect line protruding from the disk bottom edge ($t=500\,\Delta x^2/\Gamma\L$), and eventually they merge ($t=1500\,\Delta x^2/\Gamma\L$). The newly formed defect line then gets pinned to the top edge ($t=2500\,\Delta x^2/\Gamma\L$), forming two line segments that gradually move away from each other, leading to a stable quadrupolar director field configuration.}
\label{SI_fig2}
\end{figure}

\begin{figure}[h]
\centering
\includegraphics[width=\textwidth]{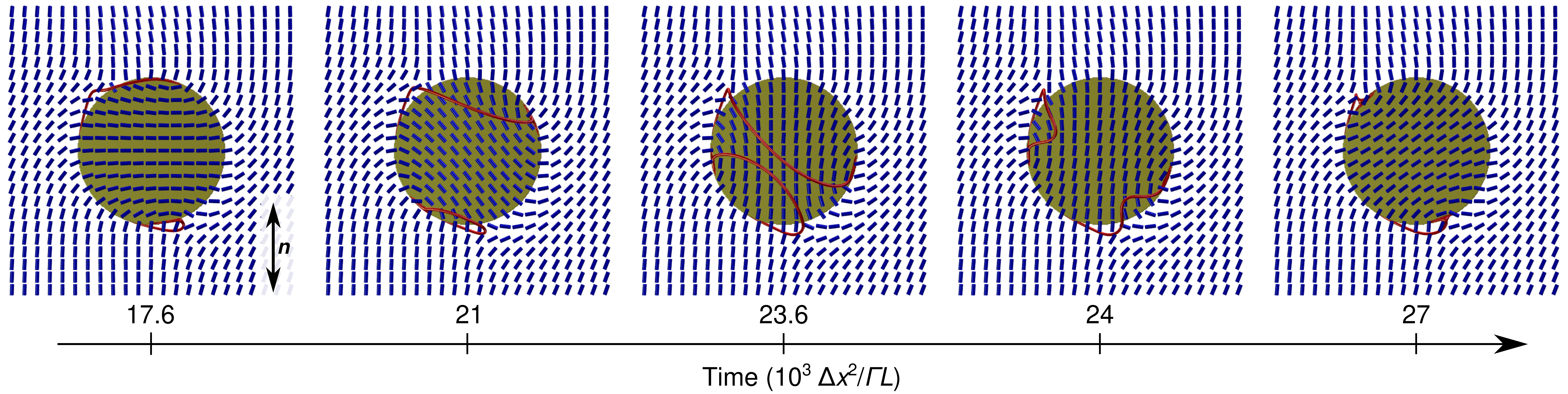}
\caption{
Defect sweeping motion in simulation. The timeline shows a period of defect motion as seen in MovieS15. Disk is viewed from the bottom directly towards the planar surface. Surface anchoring profile rotates with the disk and generates a sweeping motion of the defect lines across the disk surface.
}
\label{SI_fig3}
\end{figure}

\begin{figure}[h]
\centering
\includegraphics[width=\textwidth]{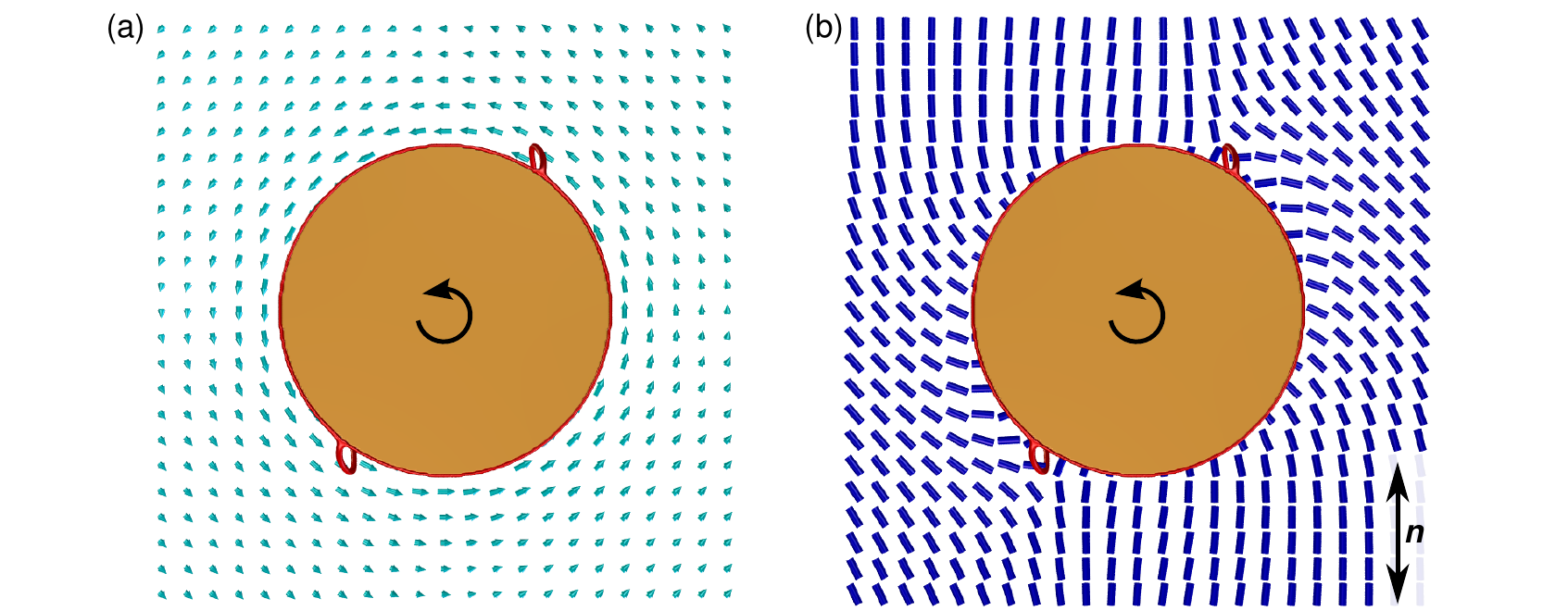}
\caption{Defect displacement in quadrupolar configuration due to rotation of the disk. 
(a) Flow field due to disk rotation. The magnitude of the flow corresponds to $\text{Er}=1.22$.
(b) Director field is distorted by the flow. Anchoring on the top and bottom plate is along the vertical direction. Rotation of the disk displaces two defect lines along the flow field. The dynamics of the transformation is shown in MovieS16.
}
\label{SI_fig4}
\end{figure}

Using the numerical method outlined above, we performed numerical simulations testing the stability and dynamics of dipolar and quadrupolar nematic solutions around a disk particle. In Figure~\ref{SI_fig1}, we test how different director profiles on the disk's side-wall affect the dipolar solution. Higher disk aspect ratios compared to experiments are used to promote the stability of the dipolar structure, since it is known that the stability of a dipolar solution is increased for large colloidal particles compared to the nematic correlation length~\cite{StarkH_PhysRep351_2001}. Since particle size is limited in simulations due to computational constraints, the dipolar solution is only metastable for specific side-wall director profiles. When the director field on the side wall is taken to be completely perpendicular to the surface, the dipolar solution is unstable and in time transforms into a quadrupolar field (Fig.~\ref{SI_fig2}). In Figure~\ref{SI_fig3}, we show a sweeping motion of a defect line underneath the disk for slow director relaxation rates at the surface. Figure~\ref{SI_fig4} shows how defect lines in a quadrupolar field deform when the disk starts to rotate.




\section{Supporting Videos}

Beside the information mentioned above, there are also 16 videos described below as the support of the text and figures.

\textbf{MovieS1}. Topological transition of quadrupolar defect to dipolar defect around a disk colloid under a counter clockwise rotating external field of period $T$=\SI{6}{\second}. The black double headed arrow indicates the far-field director. The video is 20 times faster than real time and the scale bar is \SI{50}{\micro\meter}.

\textbf{MovieS2}. Fluorescent confocal polarizing microscopy (FCPM) scanning along the z-axis of a disk colloids with dipolar defect in a planar cell filled with 5CB doped with 0.01 wt\% BTBP. The scanning depth is \SI{70}{\micro\meter} and the step length $\Delta z$ is \SI{1}{\micro\meter}. The two double headed arrows indicate the far-field director and the polarization of the excitation light. The video shows the scanning from top to bottom with a frame rate of 10 and the scale bar is \SI{50}{\micro\meter}. 

\textbf{MovieS3}. Rotation of a disk colloid in the nematic phase of 5CB with the swim stroke sweeping across the surface of the swimmer under a counter clockwise rotating external field of period $T$=\SI{40}{\second}. The black double headed arrow indicates the far-field director. The video is 10 times faster than real time and the scale bar is \SI{50}{\micro\meter}. The corresponding process is illustrated in Figs. 1(d) and 1(f) in the main text.

\textbf{MovieS4}. Trajectory planning of two disk swimmers following V-shaped trajectories under a rotating external field of period $T$=\SI{6}{\second}. The sense of rotation of the field is switched from counter clockwise to clockwise at $t$= \SI{9.2}{second} in the video. The black double headed arrow indicates the far-field director. The video is 20 times faster than real time and the scale bar is \SI{100}{\micro\meter}. The corresponding process is illustrated in Fig. 2(c) in the main text.

\textbf{MovieS5}. Trajectory planning of a disk swimmer following a curved trajectory under a counter clockwise rotating external field with changing periods from $T_1$=\SI{4}{\second} to $T_2$=\SI{12}{\second} to $T_3$=\SI{36}{\second} at $t$= \SI{1}{\second} and $t$= \SI{4}{\second} in the video. The black double headed arrow indicates the far-field director. The video is 40 times faster than real time and the scale bar is \SI{100}{\micro\meter}. The corresponding process is illustrated in Fig. 2(d) in the main text.

\textbf{MovieS6}. Rotation of a disk colloid in the isotropic phase of 5CB under a rotating external field of period $T$=\SI{20}{\second}. The video is 10 times faster than real time and the scale bar is \SI{50}{\micro\meter}.

\textbf{MovieS7}. Rotation of a $2a$=\SI{8.74}{\micro\meter} ferromagnetic spherical colloid with hedgehog defect in a planar cell filled with 5CB under a counter clockwise rotating external field of period $T$=\SI{6}{\second}. The black double headed arrow indicates the far-field director. The video is 10 times faster than real time and the scale bar is \SI{10}{\micro\meter}.

\textbf{MovieS8}. Rotation of a disk colloid with quadrupolar defect in the nematic phase of 5CB under a counter clockwise rotating external field of period $T$=\SI{60}{\second}. The black double headed arrow indicates the far-field director. The defect traveled along the edge of the disk and no translation was observed during rotation due to the lack of broken symmetry. The video is 20 times faster than real time and the scale bar is \SI{50}{\micro\meter}. 

\textbf{MovieS9}. Co-rotation of a dimer formed by two disk colloids with opposite defect polarity under a clockwise rotating external field of period $T$=\SI{80}{\second}. The black double headed arrow indicates the far-field director. The video is 40 times faster than real time and the scale bar is \SI{50}{\micro\meter}. The corresponding process is illustrated in the top row of Fig. 3(c) in the main text.

\textbf{MovieS10}. Co-rotation of a dimer formed by two disk colloids with opposite defect polarity under a clockwise rotating external field of period $T$=\SI{40}{\second}. The black double headed arrow indicates the far-field director. The video is 10 times faster than real time and the scale bar is \SI{50}{\micro\meter}. The corresponding process is illustrated in the middle row of Fig. 3(c) in the main text.

\textbf{MovieS11}. Co-rotation of a dimer formed by two disk colloids with opposite defect polarity under a clockwise rotating external field of period $T$=\SI{20}{\second}. The black double headed arrow indicates the far-field director. The video is 10 times faster than real time and the scale bar is \SI{50}{\micro\meter}. The corresponding process is illustrated in the bottom row of Fig. 3(c) in the main text.

\textbf{MovieS12}. Co-migration of two disk colloids with the same defect polarity under a counterclockwise rotating external field of period $T$=\SI{6}{\second}. The black double headed arrow indicates the far-field director. The video is 20 times faster than real time and the scale bar is \SI{50}{\micro\meter}.

\textbf{MovieS13}. Co-rotation and translation of a dimer formed by two disk colloids with the same defect polarity under a counterclockwise rotating external field of period $T$=\SI{4}{\second}. The black double headed arrow indicates the far-field director. The video is 20 times faster than real time and the scale bar is \SI{50}{\micro\meter}. The corresponding process is illustrated in Fig. 3(d) in the main text.

\textbf{MovieS14}. Dipole to quadrupole transition in a numerical simulation. Point defect in the dipolar initial structure approaches the disk and forms two disclination lines, each pinned at the bottom and at the top edge of the disk. Disclination lines move along the edges of the disk until they finally form a quadrupolar structure. The duration of the movie is $1.8\cdot 10^3\,\Delta x^2/(\Gamma L)$ in simulation units.

\textbf{MovieS15} Defect sweeping motion of a dipolar structure and a rotating disk. In the simulation, the nematic rotational viscosity parameter $\Gamma_{\text{surf}}$ is reduced by a factor of $10^4$ at the bottom surface. This causes a frustrated director field when the bottom disk surface is rotated for more than $\pi/2$ relative to the fixed director field at the bottom plane of the simulation box. The frustration is relieved by a sweeping motion of the disclination line across the disk bottom surface. Notice that the sweeping motion occurs every $\pi$ turn rotation of the bottom surface. In experiments, pinning of defect lines on the defect edge leads to a sweeping motion on every $2\pi$ turn. Simulation is performed at $Er=1.22$ and lasts for $5\cdot 10^3\,\Delta x^2/(\Gamma L)$.

\textbf{MovieS16} Quadrupolar nematic structure deformed by the flow field of the rotating disk. Two defect lines are initially at the far left and far right edge of the disk. As the disk starts to rotate, defect lines move along the flow field until they reach a static position in which the elastic force on the defect lines counterbalances the force due to the velocity field. The simulation was performed at $Er=1.22$, final structure of the velocity and the director field is shown also in Fig.~\ref{SI_fig3}.

\bibliography{supplement.bib}